\documentclass[12pt,english]{revtex4}
\usepackage[T1]{fontenc}
\usepackage[latin1]{inputenc}
\usepackage{fancyhdr}
\usepackage{array}
\usepackage{float}
\usepackage{amsmath}
\usepackage{amssymb}

\makeatletter

\usepackage{graphicx}

\usepackage{babel}
\makeatother
\begin{document}

\title{The method of Gaussian weighted trajectories. \\ V.
On the 1GB procedure for polyatomic processes \\}

\author{\textit{L. Bonnet{\footnote{Corresponding author. Email: l.bonnet@ism.u-bordeaux1.fr}}}}

\address{Institut des Sciences Moléculaires, Université Bordeaux 1,
351 Cours de la Libération, 33405 Talence Cedex, France}

\author{\textit{J. Espinosa-Garcia}}

\address{Departamento de Química Física, Universidad de Extremadura, 06071 Badajoz Spain.}

\begin{abstract}

In recent years, many chemical reactions have been studied by means of the quasi-classical trajectory (QCT) method
within the Gaussian binning (GB) procedure. The latter consists in "quantizing" the final vibrational actions in Bohr spirit by
putting strong emphasis on the trajectories reaching the products with vibrational actions close to integer values.
A major drawback of this procedure is that if $N$ is the number of product vibrational modes,
the amount of trajectories necessary to converge the calculations is $\sim$ $10^{N}$ larger than with the standard QCT method.
Applying it to polyatomic processes is thus problematic. 
In a recent paper, however, Czak\'o and Bowman propose to quantize the total vibrational energy
instead of the vibrational actions [G. Czak\'o and J. M. Bowman, J. Chem. Phys., 131, 244302 (2009)], a procedure 
called 1GB here. The calculations are
then only $\sim$ 10 times more time-consuming than with the standard QCT method, allowing thereby for considerable numerical saving. 
In this paper, we propose some theoretical arguments supporting the 1GB procedure and check its validity
on model test cases as well as the prototype four-atom reaction OH+D$_2$ $\longrightarrow$ HOD+D.

\end{abstract}

\maketitle

\section{Introduction}

Improving our ability to accurately describe gas-phase chemical reactions and inelastic collisions is 
a stimulating theoretical issue at the interface of physics and chemistry \cite{Levine} and a necessary step towards a deep understanding 
of the evolution of planetary atmospheres and interstellar clouds.

Assuming that for a given process, the electronic problem has been solved \cite{Anybook}, i.e., the potential energy of interaction between
nuclei is known, nuclear motions can be studied either quantum \cite{Launay, Enzo, Nyman, Althorpe, Schatz, Kup, Han, Enzo2} or 
classical mechanically \cite{Porter, Sewell}. For the present time, however, quantum scattering approaches can hardly be applied to more 
than three-atom processes, despite current computer performances and a great deal of methodological effort made to go beyond 
the triatomic problem \cite{Clary, Enzo3, Aron, Schmatz,Dong-Hui}.

On the other hand, the classical approach, well known as the quasi-classical trajectory 
method (QCTM) \cite{Porter, Sewell}, is much less time consuming and can therefore be applied to almost any process, 
independently on the number of atoms involved. We focus our attention on this method in the present paper.

A major goal of QCTM is to predict the distributions of the translational energy between
bimolecular collision or photodissociation products as well as the distribution of their quantum states \cite{Levine}. 
These distributions, measured in molecular beam experiments, are among the most refined data on chemical reactivity and 
molecular reaction dynamics. In this work, we concentrate on the possible descriptions of these two quantities
within QCTM. 
 
In its standard implementation, QCTM deals with the \emph{standard binning} (SB) procedure (or histogram method) 
for assigning trajectories to the various quantum 
states available. In order to introduce this procedure, we consider the three-atom exchange reaction of the type 
A + BC $\longrightarrow$ AB + C. If at the end of a given reactive trajectory, the vibrational action of AB is $x$ in units of $h$ (see 
appendix A for the mathematical definition of $x$)
and its rotational angular momentum is $j$ in units of $\hbar$, the trajectory is assumed to only contribute to the 
AB quantum state $(\bar{x},\bar{j})$ where $\bar{x}$ and $\bar{j}$ are the nearest integers of $x$ and $j$ respectively
(in the following, the nearest integer of any variable will also be denoted by the variable with a bar on top of it).

About ten years ago, however, it was suggested
that such a procedure might lead to wrong predictions when the energy available to the
products is too low for the quantum and classical densities of product states to be equal, or equivalently,
when the available quantum states are widely spaced as compared to the energy disposal \cite{BR1}. 
A \emph{Gaussian Binning} (GB) procedure was then proposed \cite{BR1} which amounts to assigning to each trajectory a Gaussian
statistical weight such that the closer the final actions to their nearest integers, the larger the weight
(by action, we mean here both vibrational actions and rotational angular momenta in the previously defined units). 
For the previous triatomic process, the Gaussian weight of the trajectory ending with $(x,j)$ is
\\
\begin{equation}
G(x,j) = G(x-\bar{x}) G(j-\bar{j})
\label{1}
\end{equation}
with
\begin{equation}
G(u) = \frac{e^{-u^2/\epsilon^2}}{\pi^{1/2} \epsilon},
\label{2}
\end{equation}
\\
$\epsilon$ being usually kept at $\sim$ 0.05 \cite{Aoiz, Bowman, BR3}. Like in the SB procedure, trajectories 
do only contribute to the quantum state defined by the center $(\bar{x},\bar{j})$ of the bin or unit square in which 
$(x,j)$ stands. The GB procedure is therefore a practical way of taking into account Bohr quantization in the analysis
of the final results. The GB procedure turns out to be a reminiscence of the use of narrow boxes proposed by Ron \emph{et al}
in the early 80's \cite{Ron}, a method apparently ignored or forgotten by QCTM users. 

Initially proposed on the basis of intuitive arguments, the GB procedure was later shown to be a
practical implementation of classical S matrix theory (CSMT) in the random phase approximation \cite{BR2, B1}, 
CSMT being the first and simplest (or least complex) semi-classical approach of molecular collisions 
pioneered by Miller and Marcus in the early seventies \cite{Miller1, Marcus, Miller2, Miller3, Child, Connor, B2}.


The Gaussian weight $G(u)$ is characterized by a full width at half maximum of $\sim$ 10 percent. This means that
the values of $x$ and $j$ respectively in the ranges $[\bar{x}-0.05,\bar{x}+0.05]$ and $[\bar{j}-0.05,\bar{j}+0.05]$
mostly contribute to the GB population of the level $(\bar{x},\bar{j})$, as compared with the values in the unit ranges 
$[\bar{x}-0.5,\bar{x}+0.5]$ and $[\bar{j}-0.5,\bar{j}+0.5]$ which contribute to the SB population. 
Therefore, the area in the $(x,j)$ plane contributing to the GB population is $\sim$ 100 times smaller than the one 
contributing to the SB population and it is necessary to run $\sim$ 100 times more trajectories within the GB procedure
than within the SB one for the same level of convergence of the final results. 

In many experiments, however, the number of available rotational states of AB is significantly larger than the number
of its vibrational states (more than $\sim$ 10 against less than $\sim$ 3) and one arrives at the same result when weighting 
the trajectories by Eq. \eqref{1} or by $G(x-\bar{x})$ alone. 
Within this partial GB procedure, corresponding to Eqs. (13) and (14) of reference 24, it is thus sufficient to run $\sim$ 10 
times more trajectories than within the HB one \cite{Aoiz, Bowman, BR3, B1, Lendv1, Lendv2, Halvick, Guo}. 

However, considering polyatomic reactions where the number of vibrational modes is easily ten or more, strongly clouds the 
situation. The reason is that "quantizing" $N$ modes amounts to weight the trajectories by a product of $N$
Gaussians. Therefore, one is led to run $\sim 10^N$ times more trajectories within the GB procedure
than within the SB one. For the reaction F+CH$_4 \longrightarrow$ FH+CH$_3$ and its isotopic variants, 
much studied experimentally in the recent years \cite{Liu}, the previous number is $\sim$ 10 millions ! 
Since one needs at least a few hundreds of thousands 
of trajectories within the HB procedure, one should run a few trillions of trajectories within the GB procedure, which 
is just not feasible. 

In order to circumvent this difficulty, Czak\'o and Bowman recently proposed to weight the trajectories by $G(u)$
(see Eq. \eqref{2}) with 
\\
\begin{equation}
u = \frac{\sum_{i=1}^N \omega_i(x_i-\bar{x}_i)}{\sum_{i=1}^N \omega_i},
\label{3}
\end{equation}
\\
$x_i$ being the vibrational action for the $i^{th}$ mode and 
$\omega_i$ the corresponding frequency \cite{Cza}. In other words, they proposed to
quantize, with one Gaussian only, the total vibrational energy (in the harmonic approximation) instead of the vibrational actions. 
Consequently, this 1GB procedure allows for a huge amount of computational savings for large systems. 

The goal of the present paper is to propose theoretical arguments supporting this procedure and check its validity
on model as well as actual processes.

The paper is organized as follows. In section II, the 1GB procedure is shown to be equivalent to the usual GB procedure for statistical
collinear processes. We then discuss the conditions for its validity in the general case.
The predictions to which it leads are compared in section III with the usual SB and GB predictions for a model test case involving
three vibrational modes. In section IV, the approach is applied to the prototype four-body chemical reaction 
OH+D$_2 \longrightarrow$ HOD+D which is among the simplest polyatomic bimolecular reactions 
\cite{Alagia,Clary2,Davis,Troya,Saracibar,Sierra,Joaquin}. 
We finally conclude in section V.


\section{Theoretical analysis of the 1GB procedure}

In a first step, we focus our attention on collinear processes in the course of which 
nuclei keep on a line fixed in the laboratory frame. The realistic three-dimensional case where rotation motions are active is
considered in a second step.

\subsection{Collisional system involving two vibrational modes}

Consider the collinear inelastic collision between atom A and the triatomic molecule BCD at the classically available energy
$E$ with respect to the free fragments. Assuming that the harmonic approximation is valid for the intra-molecular motion of BCD,
its vibrational energy $E_V$ at the end of the collision reads (see appendix A)
\\
\begin{equation}
E_V = \omega_1 (x_1+\frac{1}{2})+\omega_2 (x_2+\frac{1}{2})
\label{7}
\end{equation}
\\
where $\omega_1$ and $\omega_2$ are the energy spacings between neighboring states for the two vibrational streching 
normal modes of BCD and $x_1$ and $x_2$ are their related actions
(since B, C and D are aligned, the usual bending vibration is ignored). 

The relative translational energy $E_T$ between A and BCD satisfies the identity
\\
\begin{equation}
E_T=E-E_V.
\label{8}
\end{equation}
\\
We call $\rho(x_1,x_2)$ the classical distribution of the actions $x_1$ and $x_2$, supposed to be normalized to unity.
\\
\\
Additional paragraph 1:
\\
\\
We shall consider the formal expressions of both the translational energy distribution of the final products and the one of their
quantum states. However, we shall only represent the former distribution in the figures. We might have done the contrary, but the
translational energy distribution is by far the most widely measured in molecular beam experiments. We thus believe that discussing
the different ways this distribution can be represented in QCT studies is an important issue.
In addition to that, the translational and internal energies being mathematically related (see Eq. \eqref{8}), the two distributions
can, in principle, be deduced from each other. In this section and the next one, for instance, it will turn out that
the product state distribution is readily obtained from visual inspection of the translational energy distribution.
\\
\\
End of the additional paragraph 1.
\\
\\

\subsection{Purely classical translational energy distribution}

The translational energy distribution obtained from a strict application of classical mechanics reads 
\\
\begin{equation}
P_C(E_T) = \int\;dx_1\;dx_2\;\rho(x_1,x_2)\;\delta\bigl(E_T-E+\omega_1 (x_1+\frac{1}{2})+\omega_2 (x_2+\frac{1}{2})\bigr)
\label{9}
\end{equation}
\\
(see appendix B for its derivation). Since this density has no quantum attribute, it is usually in bad agreement with
quantum scattering and/or highly resolved experimental distributions, unless $E$ is much larger than the average quantum level 
spacing. 
\\
\\
Additional paragraph 2:
\\
\\
Nevertheless, this distribution has been so widely used in QCT studies that for the sake of completeness, we shall be 
considering it in this work. 
\\
\\
End of the additional paragraph 2.
\\
\\

\subsection{SB distributions}

A variant of the previous distribution, incorporating to some extent the idea of vibrational quantization, is as follows:
\\
\begin{equation}
P_{SB}(E_T) = \int\;dx_1\;dx_2\;\rho(x_1,x_2)\;\delta(E_T-E+\omega_1 (\bar{x}_1+\frac{1}{2})+\omega_2 (\bar{x}_2+\frac{1}{2})).
\label{10}
\end{equation}
\\
We note that the only difference with respect to Eq. \eqref{9} is that the $x_i$'s in the delta function have been replaced 
by the $\bar{x}_i$'s.

As in the SB procedure, the bins are one unit wide, the domain of integration in Eq. \eqref{10} consists of the domains
corresponding to each pair of the integer quantum numbers $n_1$ and $n_2$. Decomposing the integral in \eqref{10} into
a sum of integrals over these unit-sized domains, one gets: 
\\
\begin{equation}
P_{SB}(E_T) = \sum_{n_1,n_2}\;\int_{D_{n_1n_2}} dx_1\;dx_2\;
\rho(x_1,x_2)\;\delta(E_T-E+\omega_1 (\bar{x}_1+\frac{1}{2})+\omega_2 (\bar{x}_2+\frac{1}{2}))
\label{11}
\end{equation}
\\
where $D_{n_1n_2}$ is the unit square in the plane $(x_1,x_2)$ centered on $(n_1,n_2)$. The $\bar{x}_i$'s
being equal to the $n_i$'s in $D_{n_1n_2}$, we then arrive at
\\
\begin{equation}
P_{SB}(E_T) = \sum_{n_1,n_2}\;P_{SB}(n_1,n_2)\;\delta(E_T-E+\omega_1 (n_1+\frac{1}{2})+\omega_2 (n_2+\frac{1}{2}))
\label{12}
\end{equation}
where 
\begin{equation}
P_{SB}(n_1,n_2) = \int_{D_{n_1n_2}} dx_1\;dx_2\;\rho(x_1,x_2)
\label{13}
\end{equation}
\\
is recognized to be the SB population of the quantum state $(n_1,n_2)$.

\subsection{GB distribution}

The GB distribution is readily found from Eqs. \eqref{12} and \eqref{13} to be given by
\\
\begin{equation}
P_{GB}(E_T) = \sum_{n_1,n_2}\;P_{GB}(n_1,n_2)\;\delta(E_T-E+\omega_1 (n_1+\frac{1}{2})+\omega_2 (n_2+\frac{1}{2}))
\label{14}
\end{equation}
and
\begin{equation}
P_{GB}(n_1,n_2) = \int_{D_{n_1n_2}} dx_1\;dx_2\;G(x_1,x_2)\;\rho(x_1,x_2)
\label{15}
\end{equation}
with 
\begin{equation}
G(x_1,x_2)=G(x_1-n_1)G(x_2-n_2).
\label{15a}
\end{equation}
\\
The even unit weight in the integrand of Eq. \eqref{13} has thus been replaced by the Gaussian weight $G(x_1,x_2)$.

When making $\epsilon$ tend to zero, we arrive at a distribution which we shall call "exact" in the following.
It is of course not exact in the true quantum mechanical sense, but it is the best distribution we can arrive at by 
simple inclusion of Bohr quantization in QCTM. In this limit,
$G(u)$ tends to the delta-function $\delta(u)$ and Eq. \eqref{15} reads
\\
\begin{equation}
P_{E}(n_1,n_2) = \int_{D_{n_1n_2}} dx_1\;dx_2\;\delta(x_1-n_1)\;\delta(x_2-n_2)\;\rho(x_1,x_2),
\label{16}
\end{equation}
\\
giving immediately
\begin{equation}
P_{E}(n_1,n_2) = \rho(n_1,n_2).
\label{16a}
\end{equation}
Then, Eq. \eqref{14} reads
\\
\begin{equation}
P_{E}(E_T) = \sum_{n_1,n_2}\;\rho(n_1,n_2)\;\delta(E_T-E+\omega_1 (n_1+\frac{1}{2})+\omega_2 (n_2+\frac{1}{2})).
\label{17}
\end{equation}
\\
Since the 1GB distribution to be derived in the next section is supposed to be an alternative to the GB one, the former will be
systematically tested against the latter and its "exact" limit in the followings.



\subsection{1GB distributions}

\subsubsection{Statistical case}

\begin{figure}[H]
\begin{center}\includegraphics[%
  clip,
  angle=90,
  origin=c]{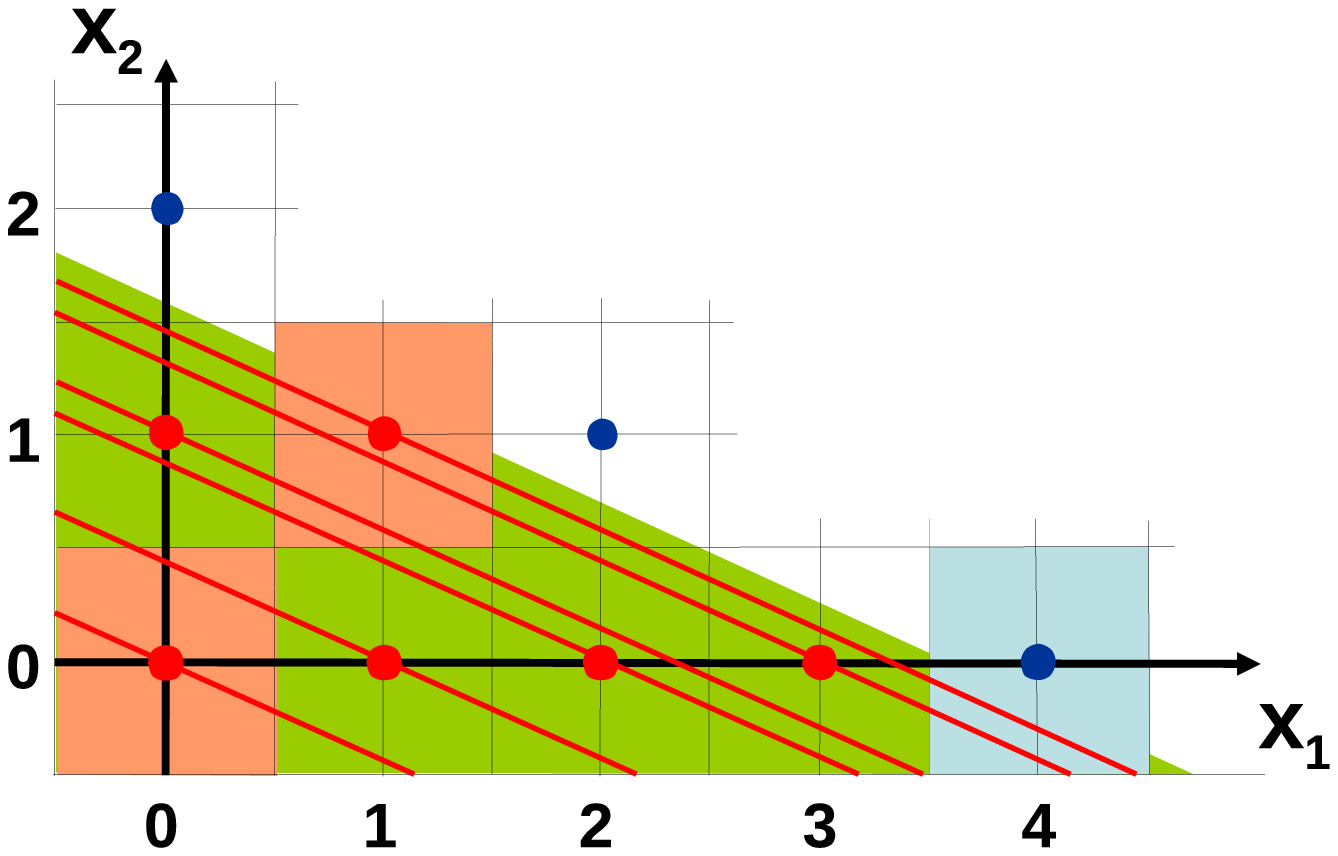}\end{center}

\caption{\label{fig:plot1}}
\end{figure}

Indirect chemical reactions involving long-lived complexes have been the subject of intense research during 
the last few years \cite{BR6, BR4, Tomas1, Tomas2, Tomas3, BR5}. 
\\
\\
Additional paragraph 3:
\\
\\
In Phase Space Theory, the simplest statistical
approach (see references \cite{BR4} and \cite{BR5} and references therein), 
the final product states consistent with total energy and total angular momentum are equally probable.
\\
\\
End of the additional paragraph 3.
\\
\\
For the present system, the analogous situation corresponds to a uniform density $\rho(x_1,x_2)$ in the energetically 
available action space defined by
\\
\begin{equation}
E \ge \omega_1 (x_1+\frac{1}{2})+\omega_2 (x_2+\frac{1}{2})
\label{20}
\end{equation}
\\
and both $x_1$ and $x_2$ greater than minus 1/2. This triangular domain is represented in Fig.~\ref{fig:plot1} 
for $E$, $\omega_1$ and $\omega_2$ kept at 5.2, 1 and 2.3 respectively ; these values have been chosen in such 
a way that six quantum states, indicated by red dots, are available (different values might have been chosen as well). 
Three forbidden quantum states are represented by dark blue dots and
three unit squares centered on quantum states are emphasized. The two salmon ones
correspond to available quantum states while the light blue one corresponds to a forbidden state.

Throughout the present part, $\rho(x_1,x_2)$ will be simply denoted $\rho$. Its value is 
\\
\begin{equation}
\rho = \frac{2\omega_1 \omega_2}{E^2}
\label{20a}
\end{equation}
\\
(the inverse of the area of the green triangle) inside the triangle and zero outside. 

$P_C(E_T)$ can be determined analytically by using the identity
\\
\begin{equation}
\delta(a x) = \frac{1}{|a|}\delta(x),
\label{20c}
\end{equation}
leading to
\begin{equation}
P_C(E_T) = \frac{2}{E}\Big(1-\frac{E_T}{E}\Big).
\label{20d}
\end{equation}
\\
\\
\\
Additional paragraph 4:
\\
\\
In principle, the translational energy distribution measured in a perfect experiment would consist in 
a set of Dirac delta functions for the energies complementary with those of the allowed product quantum
states. However, no experiment is "perfect". There is always an uncertainty in both the total energy 
available to the products and the measure of the translational energy. Consequently, the peaks are 
necessarily broaden.
In order to take into account
this uncertainty in the theory, we shall replace the Dirac peaks present in Eqs. \eqref{12}, \eqref{14} and \eqref{17} 
by $G(u)$ (see Eq. \eqref{2}) with $\epsilon = 0.05$, meaning that the uncertainty on $E_T$ is $\sim$ 0.1. 
This arbitrary value of $\epsilon$ makes the peaks neither too narrow, nor too broad as compared to the total 
energy $E$. The fact that it was kept at the same value as in the GB procedure should not confuse the reader.  
Its choice was a matter of convenience, nothing else.

It is worth emphasizing that there are thus two separate issues in this work: the first, and central one, is the use of Gaussians
to deal with Bohr quantization in QCT calculations ; the second, and minor one, is the use of Gaussians
to take into account the uncertainty in the measurement of the translational energy. 
\\
\\
End of the additional paragraph 4.
\\
\\
$P_{SB}(E_T)$ and $P_{GB}(E_T)$ were determined by Monte-Carlo integration over $x_1$ and $x_2$,
using $N_T = 200000$ points randomly chosen in the triangular domain. 
The corresponding expressions are
\\
\begin{equation}
P_{X}(E_T) = \sum_{n_1,n_2}\;P_{X}(n_1,n_2)\;G(E_T-E+\omega_1 (n_1+\frac{1}{2})+\omega_2 (n_2+\frac{1}{2}))
\label{12a}
\end{equation}
\\
where $X$ stands for $SB$ or $GB$, with
\\
\begin{equation}
P_{SB}(n_1,n_2) = \frac{N_{D_{n_1n_2}}}{N_T},
\label{12a1}
\end{equation}
\\
$N_{D_{n_1n_2}}$ being the number of trajectories ending in $D_{n_1n_2}$, and
\\
\begin{equation}
P_{GB}(n_1,n_2) = \frac{1}{N_T} \sum_{k=1}^{N_{D_{n_1n_2}}}\;G(x^k_1,x^k_2),
\label{12a2}
\end{equation}
\\
$x^k_1$ and $x^k_2$ being the final actions for the $k^{th}$ trajectory ending in $D_{n_1n_2}$.

\begin{figure}[H]
\begin{center}\includegraphics[%
  clip,
  angle=0,
  origin=c]{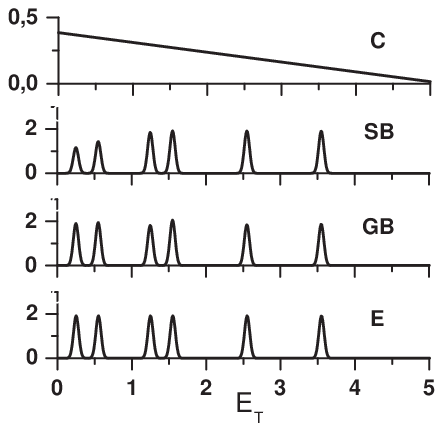}\end{center}

\caption{\label{fig:plot2}}
\end{figure}

The four distributions are represented in Fig.~\ref{fig:plot2} for 500 values of $E_t$ regularly distributed between 0 and $E$.

$P_{C}(E_T)$ appears to be in complete disagreement with the "benchmark" distribution $P_{E}(E_T)$, for by construction,
no structure can be reproduced. $P_{SB}(E_T)$ takes the structures into account, but the heights of 
the peaks are inaccurate for the small values of $E_T$. 
Conversely, $P_{GB}(E_T)$ 
is in excellent agreement with $P_{E}(E_T)$, as expected. 
Note that for these two distributions, the peaks have the same
height, meaning that the populations of the available quantum states are all equal, in agreement with the statistical
hypothesis.  

The heights of the peaks of $P_{SB}(E_T)$ increase with $E_t$. This can be easily understood from Fig.~\ref{fig:plot1}.
The values of the translational energies corresponding to the top of 
the peaks (see Fig.~\ref{fig:plot2}) are given by
\\
\begin{equation}
E_{n_1n_2}=E-\Big(\omega_1 (n_1+\frac{1}{2})+\omega_2 (n_2+\frac{1}{2})\Big),
\label{21}
\end{equation}
\\
the integers $n_1$ an $n_2$ being such that $(n_1,n_2)$ is an allowed quantum state. These states are $(1,1)$, $(3,0)$, 
$(0,1)$, $(2,0)$, $(1,0)$ and $(0,0)$, by order of increasing translational energy. 
The straight lines defined by 
\\
\begin{equation}
\omega_1 (x_1+\frac{1}{2})+\omega_2 (x_2+\frac{1}{2})=E-E_{n_1n_2}
\label{21}
\end{equation}
\\
are represented in Fig.~\ref{fig:plot1}. Clearly, the spacings between these lines exactly follow the spacings 
between the nearest $E_{n_1n_2}$'s (see Fig.~\ref{fig:plot2}). 

Now, the heights are proportional to the SB populations,
themselves proportional to the available areas of the unit squares centered on the quantum states. From Fig.~\ref{fig:plot1},
it is clear that for $(0,0)$, this area, represented in salmon, 
is 1, but for $(1,1)$, it is only equal to $\sim$ 0.6. One also guesses that for $(3,0)$ and $(0,1)$, the areas are equal to
$\sim$ 0.7 and $\sim$ 0.9 respectively while for the remaining states $(1,0)$ and $(2,0)$, they are equal to $\sim$ 1. 
This explains why the heights of the first three peaks of $P_{SB}(E_T)$ (see Fig.~\ref{fig:plot2}) 
are only $\sim$ 60, $\sim$ 70 and $\sim$ 90 percent of the height of the remaining peaks.
This cannot happen with the GB distribution, for with sufficiently thin Gaussians, their value is the same for all 
the quantum states. 

\begin{figure}[H]
\begin{center}\includegraphics[%
  clip,
  angle=0,
  origin=c]{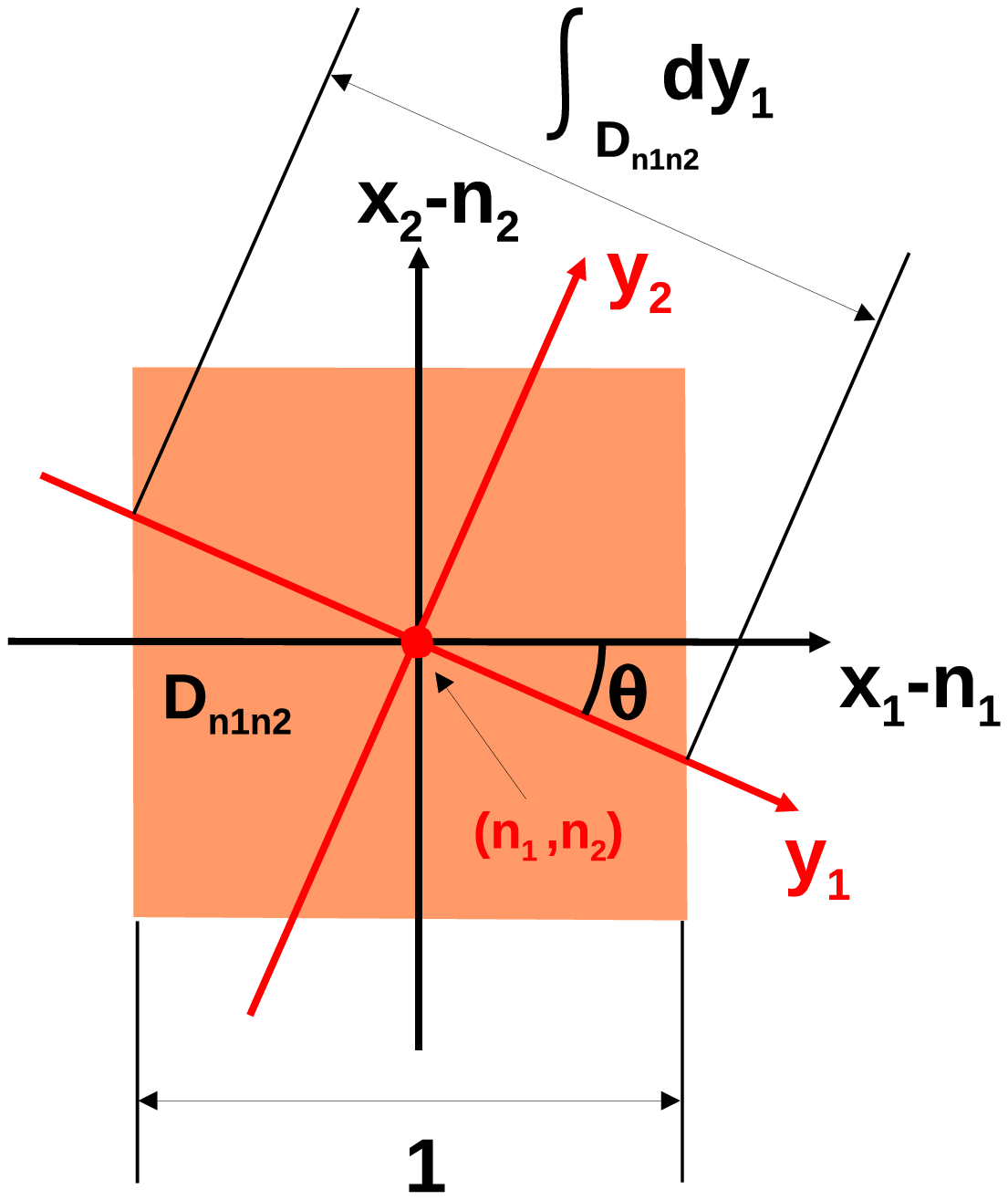}\end{center}

\caption{\label{fig:plot3}}
\end{figure}

Having defined the system of interest and applied the different methods currently utilized in QCT calculations today,
we are now in a position to introduce the 1GB procedure. $P_{E}(n_1,n_2)$, 
given by Eq. \eqref{16} with $\rho(x_1,x_2)=\rho$, can be rewritten as
\\
\begin{equation}
P_{E}(n_1,n_2) = \rho\;\int_{D_{n_1n_2}} dy_1\;dy_2\;\delta(y_1)\;\delta(y_2)
\label{22}
\end{equation}
\\
where $y_1$ and $y_2$ are two new coordinates related to $x_1$ and $x_2$ by the rotation
\\ 
\begin{equation}
y_1=cos\theta\;(x_1-n_1)-sin\theta\;(x_2-n_2)
\label{23}
\end{equation}
and
\begin{equation}
y_2=sin\theta\;(x_1-n_1)+cos\theta\;(x_2-n_2)
\label{24}
\end{equation}
with 
\begin{equation}
cos\theta=\frac{\omega_2}{(\omega_1^2+\omega_2^2)^{1/2}}
\label{25}
\end{equation}
and 
\begin{equation}
sin\theta=\frac{\omega_1}{(\omega_1^2+\omega_2^2)^{1/2}}.
\label{26}
\end{equation}
\\
The axis $y_1$ runs through $(n_1,n_2)$ and is parallel to the hypotenuse of the green triangle. $y_1$
is thus one of the red axes represented in Fig.~\ref{fig:plot1}.
The axis $y_2$ also runs through $(n_1,n_2)$ and is orthogonal to $y_1$. These axes are represented 
in Fig.~\ref{fig:plot3} as well as the unit square $D_{n_1n_2}$. It is clear that integration over 
$y_1$ and $y_2$ in Eq. \eqref{22} immediately leads to Eq. \eqref{16a}.

Let us now define the population  
\\
\begin{equation}
P'_{E}(n_1,n_2) = \alpha\;\rho\;\int_{D_{n_1n_2}} dy_1\;dy_2\;\delta(y_2)
\label{27}
\end{equation}
where
\begin{equation}
\alpha = max(cos\theta,sin\theta).
\label{27a}
\end{equation}
\\
As compared with Eq. \eqref{22}, Eq. \eqref{27} involves only one Dirac distribution. After a trivial integration
with respect to $y_2$, $P'_{E}(n_1,n_2)$ reads
\\
\begin{equation}
P'_{E}(n_1,n_2) = \alpha\;\rho\;\int_{D_{n_1n_2}} dy_1.
\label{28}
\end{equation}
\\
However, one deduces from Fig.~\ref{fig:plot3}, corresponding to the case where $cos\theta$ is greater than $sin\theta$
($\theta$ lower than $\pi/4$), that
\\
\begin{equation}
cos\theta = \frac{1}{\int_{D_{n_1n_2}} dy_1}.
\label{29}
\end{equation}
\\
In the same way, it can be easily shown that when $cos\theta$ is smaller than $sin\theta$, 
\\
\begin{equation}
sin\theta = \frac{1}{\int_{D_{n_1n_2}} dy_1}.
\label{29a}
\end{equation}
\\
$\alpha$, given by Eq. \eqref{27a}, can thus be rewritten as 
\\
\begin{equation}
\alpha = \frac{1}{\int_{D_{n_1n_2}} dy_1},
\label{29a1}
\end{equation}
and Eq. \eqref{28} leads to
\begin{equation}
P'_{E}(n_1,n_2) = \rho.
\label{29b}
\end{equation}
\\
We thus arrive at the conclusion that the $P'_{E}(n_1,n_2)$'s are equal to the $P_{E}(n_1,n_2)$'s (see Eq. \eqref{16a}). 
Consequently, $P_{E}(n_1,n_2)$ can be rewritten from Eqs. \eqref{24}-\eqref{27} as
\\
\begin{equation}
P_{E}(n_1,n_2) = \alpha\;\int_{D_{n_1n_2}} dy_1\;dy_2\;
\delta\Big(\frac{\omega_1\;(x_1-n_1)+\omega_2\;\;(x_2-n_2)}{(\omega_1^2+\omega_2^2)^{1/2}}\Big)\;\rho.
\label{29c}
\end{equation}
\\
Moreover, from Eqs. \eqref{25}, \eqref{26} and \eqref{27a} and using the fact that $\alpha \delta(q) = \delta(q/\alpha)$ 
(deduced from Eq. \eqref{20c}), we obtain
\\
\begin{equation}
P_{E}(n_1,n_2) = \int_{D_{n_1n_2}} dy_1\;dy_2\;
\delta\Big(\frac{\omega_1\;(x_1-n_1)+\omega_2\;\;(x_2-n_2)}{max(\omega_1,\omega_2)}\Big)\;\rho.
\label{29c1}
\end{equation}
\\
Applying the GB procedure, i.e. replacing $\delta$ by $G$, and going back to the $x_i$'s, $P_{E}(n_1,n_2)$ 
finally reads
\\
\begin{equation}
P_{1GB}(n_1,n_2) = \int_{D_{n_1n_2}} dx_1\;dx_2\;
G\Big(\frac{\omega_1\;(x_1-n_1)+\omega_2\;\;(x_2-n_2)}{max(\omega_1,\omega_2)}\Big)\;\rho(x_1,x_2),
\label{29d}
\end{equation}
\\
an expression formally close to the expression
\\
\begin{equation}
P_{1GB}(n_1,n_2) = \int_{D_{n_1n_2}} dx_1\;dx_2\;
G\Big(\frac{\omega_1\;(x_1-n_1)+\omega_2\;\;(x_2-n_2)}{\omega_1+\omega_2}\Big)\;\rho(x_1,x_2)
\label{29e}
\end{equation}
\\
corresponding to the calculations of Czak\'o and Bowman \cite{Cza}. 

\begin{figure}[H]
\begin{center}\includegraphics[%
  clip,
  angle=0,
  origin=c]{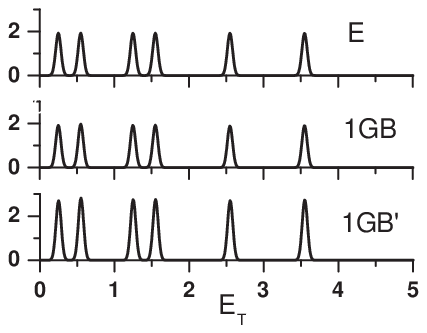}\end{center}

\caption{\label{fig:plot4}}
\end{figure}

$P_{1GB}(E_t)$ is then deduced from $P_{1GB}(n_1,n_2)$ by
\\
\begin{equation}
P_{1GB}(E_T) = \sum_{n_1,n_2}\;P_{1GB}(n_1, n_2)\;\delta(E_T-E+\omega_1 (n_1+\frac{1}{2})+\omega_2 (n_2+\frac{1}{2})).
\label{29f}
\end{equation}
\\
Like previously, the $\delta$ function in the above expression is replaced by a Gaussian in order to take into account 
the uncertainty in the measurement of the translational energy. Moreover, the Monte-Carlo expression of $P_{1GB}(n_1, n_2)$ reads
\\
\begin{equation}
P_{1GB}(n_1,n_2) = \frac{1}{N_T} \sum_{k=1}^{N_{D_{n_1n_2}}}\;
G\Big(\frac{\omega_1\;(x^k_1-n_1)+\omega_2\;\;(x^k_2-n_2)}{\omega}\Big),
\label{29g}
\end{equation}
with
\begin{equation}
\omega = max(\omega_1,\omega_2)
\label{29h}
\end{equation}
according to Eq. \eqref{29d} or 
\begin{equation}
\omega = \omega_1+ \omega_2
\label{29i}
\end{equation}
\\
according to Eq. \eqref{29e}. Eqs. \eqref{2}, \eqref{29g} and \eqref{29i} correspond to 
Eqs. 13 and 16 in the paper by Czak\'o and Bowman \cite{Cza}.
The distribution obtained from Eqs. \eqref{29f}-\eqref{29h} is represented in Fig.~\ref{fig:plot4} (1GB curve), 
together with $P_E(E_T)$ (like previously, $\epsilon$ was kept at 0.05 for all the Gaussians). 
The two densities are in such a good agreement that they cannot be distinguished.
On the other hand, the distribution obtained from Eqs. \eqref{29f}, \eqref{29g} and \eqref{29i}, also shown in 
Fig.~\ref{fig:plot4} (1GB' curve), has the good shape, but its norm is too large. We shall come back to this
important normalization issue in section II.H.
\\
\\
Additional paragraph 5:
\\
\\
One may wonder wether the selection of $y_2$ to be the argument of the delta function in Eq. \eqref{27} is arbitrary.
For instance, one might be tempted by stating in view of Fig.~\ref{fig:plot3} that interchanging $y_1$ and $y_2$ still
leads to the equality between $P_{E}(n_1,n_2)$ and $P'_{E}(n_1,n_2)$. This is indeed true for the less excited states 
$(0,0)$, $(1,0)$, $(2,0)$ and $(0,1)$, but not for the most excited states $(3,0)$ and $(1,1)$ (see Fig.~\ref{fig:plot1}). As a matter
of fact, the upper limit of the green triangle would limit the integration along the new $y_1$ axis (corresponding to the $y_2$ axis
in Fig.~\ref{fig:plot3}) in such a way
that for $(3,0)$ and $(1,1)$, $P'_{E}(n_1,n_2)$ would be lower than $P_{E}(n_1,n_2)$. 
On the other hand, slightly varying $\theta$ about
the value defined by Eqs. \eqref{25} and \eqref{26} will preserve the equality between $P_{E}(n_1,n_2)$ and $P'_{E}(n_1,n_2)$ 
provided that the hypotenuse of the green triangle is not too close to the most excited states. 
Strictly speaking, the value of $\theta$ defined by Eqs. \eqref{25} and \eqref{26} is thus not the only
satisfying one. Nevertheless, it is the only one for which the equality between $P_{E}(n_1,n_2)$ and $P'_{E}(n_1,n_2)$
is systematically satisfied, no matter how close to the most excited states the hypotenuse is.
This makes it superior to any other one.
\\
\\
End of additional paragraph 5.
\\
\\ 
\subsubsection{Non statistical case}

What about non statistical situations ? Given that $\rho(x_1,x_2)$ is non uniform in the energetically allowed triangle,
Eq. \eqref{28} reads
\\
\begin{equation}
P'_{E}(n_1,n_2) = \alpha\;\int_{D_{n_1n_2}} dy_1\;\rho(x_1,x_2).
\label{27b}
\end{equation}
\\
Recall that in the statistical case, the 1GB procedure was justified by the fact that $P'_{E}(n_1,n_2)$ is
equal to $P_{E}(n_1,n_2)$.

In the present case, $P'_{E}(n_1,n_2)$ is still a reasonable approximation of $P_{E}(n_1,n_2)$ provided that 
the variation of $\rho(x_1,x_2)$ along the $y_1$-axis is sufficiently smooth. If so, it is indeed clear using Eq. \eqref{29a1} that
\\
\begin{equation}
\alpha \int_{D_{n_1n_2}} dy_1\;\rho(x_1,x_2) = \frac{\int_{D_{n_1n_2}} dy_1\;\rho(x_1,x_2)}{\int_{D_{n_1n_2}} dy_1} \approx \rho(n_1,n_2),
\label{30}
\end{equation}
\\
the strict equality occuring when $\rho(x_1,x_2)$ varies linearly along the $y_1$-axis (and is non zero within $D_{n_1n_2}$). 
From Eq. \eqref{16a}, we then arrive at the conclusion that $P'_{E}(n_1,n_2)$ is roughly equal to $P_{E}(n_1,n_2)$. 

Such a smooth variation of the density in the action space 
was observed several times in the case of three-atom exchange reactions, the only difference
being that the previously introduced $x$ and $j$ actions play the role of $x_1$ and $x_2$. Fig. 4 in reference
\cite{Bowman} is a clear illustration of this statement in the case of the direct 
reaction O($^3$P)+HCl $\longrightarrow$ OH+Cl($^2$P). 
The energetically available area in the 
$(x,j)$ plane is given by
\\
\begin{equation}
E \ge \omega (x+\frac{1}{2})+B j^2.
\label{31}
\end{equation}
\\
Its upper limit is thus found to be a curved, instead of straight, line. Moreover, the lines equivalent to the 
six straight lines in Fig.~\ref{fig:plot1} are also curved. Along these lines (not drawn in Fig. 4 of reference
\cite{Bowman}), the density $\rho(x,j)$ evolves rather smoothly. In other words, despite the fact that 
the distribution of the translational energy is not statistical, 
the intra-molecular vibrational redistribution (IVR) in the strong coupling region tends to distribute in a 
relatively democratic way the rest of the energy among the vibrational and rotational degrees-of-freedom. 

One expects a similar redistribution will also take place among the vibrational modes, due to their couplings. 
As shown in section IV, this is at least the case for OH+D$_2$ which, as O($^3$P)+HCl, is a direct process.

\subsection{Collisional system involving three vibrational modes}

We now consider the collinear inelastic collision between atom A and the tetra-atomic molecule BCDE involving three vibrational
normal modes. Following a reasoning analogous to the one developed in the previous subsection, we arrived after some steps of algebra to 
an expression exact (with a Gaussian of zero width) in the statistical limit ($P_E(E_T)$ being still considered as the "exact" density). With 
$\bold{n} = (n_1,n_2,n_3)$, $\bold{x} = (x_1,x_2,x_3)$, $D_{\bold{n}}$ the unit cube centered on $\bold{n}$, and
$\omega_k$, $\omega_l$ and $\omega_m$ deduced from any cyclic permutation
of $\omega_1$, $\omega_2$ and $\omega_3$, this expression is
\\
\begin{equation}
P_{1GB}(E_T) = \sum_{\bold{n}}\;P_{1GB}(\bold{n})\;\delta\Big(E_T-E+\sum_{i=1}^{3}\omega_i (x_i+\frac{1}{2})\Big)
\label{32}
\end{equation}
with
\begin{equation}
P_{1GB}(\bold{n}) = \int_{D_{\bold{n}}} d\bold{x}\;G(\frac{\sum_{i=1}^{3}\omega_i (x_i-n_i)}{\omega})\;\rho(\bold{x}),
\label{33}
\end{equation}
\\
$\omega$ being defined as 
\begin{equation}
\omega=\frac{\omega_k}{1-\frac{(\omega_k-\omega_l-\omega_m)^2}{4\omega_l\omega_m}}
\label{34}
\end{equation}
\\
if $\omega_k < \omega_l+\omega_m$ for the three cyclic permutations, or
\\
\begin{equation}
\omega=max(\omega_k,\omega_l,\omega_m)
\label{36}
\end{equation}
\\
if $\omega_k \ge \omega_l+\omega_m$ for only one of the permutations. 
In the second case,
the formulation (see Eqs. \eqref{33} and \eqref{36}) is a straightforward extension of Eq. \eqref{29d}. 

We shall retain from the above developments that for three vibrational modes, the formulation of $\omega$ is not unique. 
It depends on the values of the vibrational frequencies $\omega_1$, $\omega_2$ and $\omega_3$, 
contrary to the formulation for two vibrational modes. 

Hence, if one does not use the correct expression of $\omega$, one does not find the correct populations. 
However, the wrong populations turn out to be proportional to the correct ones. The proof is straightforward:
if we call $\omega_c$ the correct value of $\omega$ and $\omega_w$ the wrong one, the wrong populations
are found from Eqs. \eqref{33} (with $G$ replaced by the Dirac distribution) and \eqref{20c} to be equal to $\omega_w/\omega_c$ 
times the correct populations. This explains why the peaks of the 1GB' distribution in Fig.~\ref{fig:plot4} (given by Eqs. 
\eqref{29e} and \eqref{29f})
are $(\omega_1+\omega_2)/max(\omega_1,\omega_2)$ higher than the peaks of the 1GB distribution 
(given by Eqs. \eqref{29d} and \eqref{29f}).

We did not extend the above developments in the case of systems involving more than three vibrational modes for 
the mathematical developments became very tedious. Therefore, we do not know the analytical expression of 
$\omega$ making the 1GB distribution in close agreement with the "exact" or GB distribution in the statistical
limit. However, we go round this difficulty in section II.H.

\subsection{General collisions}

The extension of Eq. \eqref{9} to a three-dimensional collision involving $N$ vibrational modes is 
\begin{equation}
P_C(E_T) = \int\;dE_R\;P_C(E_T,E_R)
\label{9a}
\end{equation}
where
\begin{equation}
P_C(E_T,E_R) = \int\;d\bold{x}\;\rho(\bold{x},E_R)\;\delta\bigl(E_T-(E-E_R)+\sum_{i=1}^{N}\omega_i (x_i+\frac{1}{2})\bigr),
\label{9b}
\end{equation}
\\
$\bold{x} = (x_1,...,x_N)$, $E_R$ is the final product rotational energy and $\rho(\bold{x},E_R)$
is the distribution of $\bold{x}$ and $E_R$. 

For a given value of $E_R$, $P_C(E_T,E_R)$ appears to be formally identical to $P_C(E_T)$ in Eq. \eqref{9}
and consequently, all the developments following Eq. \eqref{9} could be repeated here identically.
The main conclusion of this section is thus the same as before, i.e., the 1GB procedure
leads to nearly the same conclusions as the GB procedure provided that 
the variation of $\rho(\bold{x},E_R)$ in any plane parallel to the plane
\\
\begin{equation}
E=E_R+\sum_{i=1}^{N}\omega_n (x_i+\frac{1}{2})
\label{9c}
\end{equation}
\\
is sufficiently smooth. As stated before, however, we do not know the expressions analogous to 
Eq. \eqref{29d} and Eqs. \eqref{33}-\eqref{36} for $N$ larger than 3.

\subsection{Normalization procedure in realistic calculations}

The Monte-Carlo expression of the
1GB populations of the final product quantum states $\bold{n} = (n_1,...,n_N)$ is given by
\\
\begin{equation}
P_{1GB}(\bold{n}) = \frac{1}{N_T}\sum_{k=1}^{N_{\bold{n}}}\;G\Big(\frac{\sum_{i=1}^N \omega_i(x^k_i-n_i)}{\omega}\Big)
\label{38}
\end{equation}
\\
where $N_T$ is the total number of trajectories run and
$N_{\bold{n}}$ is the number of trajectories ending in the product channel with $\bold{x} = (x_1,...,x_N)$ pointing in 
$D_{\bold{n}}$, the $N$-dimensional unit cube centered on $\bold{n}$. 

As seen before, however, we do not know in the general case the expression of $\omega$ leading to 1GB distributions in close
agreement with the GB ones in the statistical limit. Nevertheless, the former are proportional to the latter. 

One might thus think about re-normalizing 1GB distributions so as to give them the GB norms. But 
GB norms have no reason to be exactly equal to one, 
so it is preferable to directly re-normalize 1GB distributions to unity. The corresponding
expression is
\\
\begin{equation}
P_{1GB}(\bold{n}) = \frac
{\sum_{k=1}^{N_{\bold{n}}}\;G\Big(\frac{\sum_{i=1}^{N}\omega_i (x^k_i-n_i)}{\omega}\Big)}
{\sum_{k=1}^{N_T}\;G\Big(\frac{\sum_{i=1}^{N}\omega_i (x^k_i-n_i)}{\omega}\Big)}
\label{39}
\end{equation}
\\
where $\omega$ may be kept at the maximum of the frequencies or their sum. The final result
is not expected to depend significantly on this choice, for it will affect both the numerator and the 
denominator of Eq. \eqref{39} in nearly the same way. This point is illustrated in section IV in the case
of the reaction OH + D$_2$. Note that in the denominator, the sum is over the whole set
of computed trajectories, be they reactive or not. The various quantities in the argument of the Gaussian
are thus either those of the products and those of the reformed reagents.

Special care should however be taken with processes 
involving a large amount of vibrationally elastic non reactive trajectories. Ion-molecule reactions are 
a typical example. For such processes, an alternative to Eq. \eqref{39} is
\\
\begin{equation}
P_{1GB}(\bold{n}) = \frac
{min\Big[N_{\bold{n}},\sum_{k=1}^{N_{\bold{n}}}\;G\Big(\frac{\sum_{i=1}^{N}\omega_i (x^k_i-n_i)}{\omega}\Big)\Big]}
{\sum_{\bold{m}}\;min\Big[N_{\bold{m}},\sum_{k=1}^{N_{\bold{m}}}\;G\Big(\frac{\sum_{i=1}^{N}\omega_i (x^k_i-m_i)}{\omega}\Big)\Big]}
\label{40}
\end{equation}
\\
where the sum over $\bold{m}$ in the denominator involves the whole set of final vibrational states, i.e., those of the products as
well as those of the reformed reactants. Eq. \eqref{40} is a compact form of Eqs. (15) and (16) of reference 23. Eq. (4) of reference 34
may be a second alternative.

\section{Non statistical test case for three vibrational modes}

We still consider the collinear inelastic collision between atom A and the tetra-atomic molecule BCDE involving three harmonic
normal modes. $E$, $\omega_1$, $\omega_2$ and $\omega_3$ are respectively kept at 15, 1, 1.7 and 2.9.
We also consider the non statistical Gaussian density $\rho(x_1,x_2,x_3)$ given by
\\
\begin{equation}
\rho(x_1,x_2,x_3) = \Pi_{i=1}^{3}\;G(x_i-x^0_i)
\label{41}
\end{equation}
\\
with $x^0_1=2.2$,  $x^0_2=1.3$ and $x^0_3=0.7$, 
$\epsilon$ being kept at $0.8$ in $G(x_1-x^0_1)$, $1.4$ in $G(x_2-x^0_2)$ and $0.4$ in $G(x_3-x^0_3)$. 

\begin{figure}[H]
\begin{center}\includegraphics[%
  clip,
  angle=0,
  origin=c]{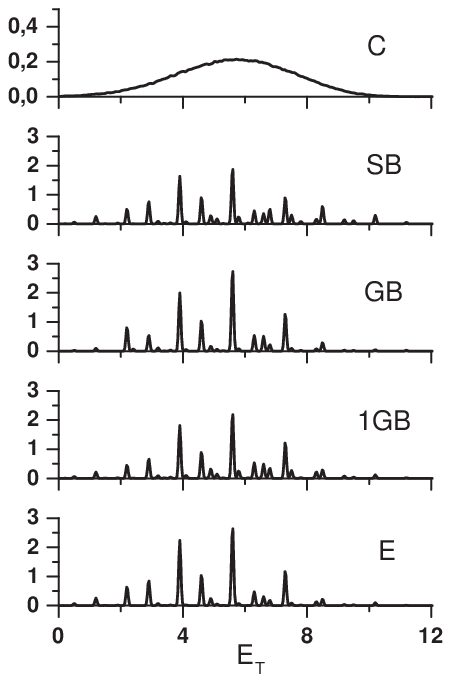}\end{center}

\caption{\label{fig:plot5}}
\end{figure}

The resulting distributions $P_{C}(E_T)$, $P_{SB}(E_T)$, $P_{GB}(E_T)$, $P_{1GB}(E_T)$ 
and $P_{E}(E_T)$, given by expressions similar to those of the previous section with one more
dimension, are represented in Fig.~\ref{fig:plot5}. The details of the calculations are exactly the same as in section II.D.1,
the only difference being that all the distributions were numerically re-normalized to unity. 

Like in the previous statistical example, $P_{C}(E_T)$ is in poor agreement with $P_{E}(E_T)$.
On the other hand, $P_{GB}(E_T)$ is in very good agreement with $P_{E}(E_T)$ ; given the large 
number of points considered in the Monte-Carlo integration, this is an expected result
despite the already "large" number of vibrational modes involved in the collision. 

Interestingly, $P_{1GB}(E_T)$ is even in slightly better accord with $P_{E}(E_T)$ than $P_{GB}(E_T)$
when looking at the details. Such a high level of agreement despite the non-statistical nature of the present process
is pleasing. It supports the statement of subsection II.E.2 that for a sufficiently smooth distribution of the final
actions, the 1GB procedure represents an accurate alternative to the GB one.

Last but not least, $P_{GB}(E_T)$ does also a good job, though
the heights of the peaks corresponding to the largest energies tend to be overestimated.  



\section{The simplest polyatomic reaction OH+D$_2$ $\longrightarrow$ HOD+D}

\begin{figure}[H]
\begin{center}\includegraphics[%
  clip,
  angle=0,
  origin=c]{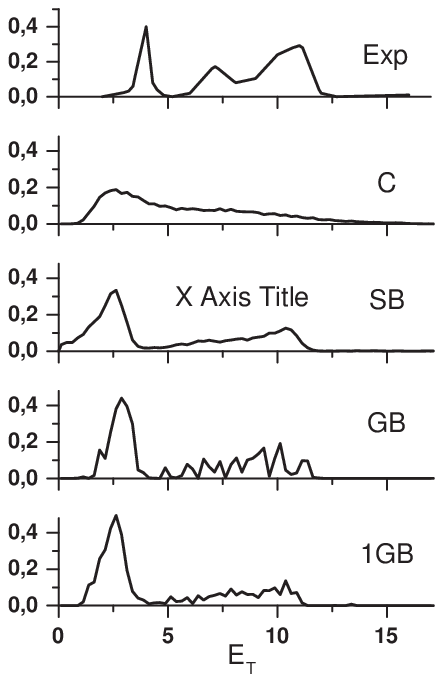}\end{center}

\caption{\label{fig:plot6}}
\end{figure}

This process has been the subject of intense research, both experimentally and theoretically 
\cite{Alagia,Clary2,Davis,Troya,Saracibar,Sierra,Joaquin,Ochoa}.
Its mechanism has been well established as being direct, with the products preferentially backward scattered, suggestive of a rebound mechanism. 
The product translational energy distribution was experimentally measured by Alagia et al. \cite{Alagia}, 
and by Davis and co-workers four years later \cite{Davis}.
While Alagia et al. found a broad and single-peaked distribution, Davis et al. found a better resolved distribution involving 
three peaks corresponding 
to the HOD vibrational states $(n_1,n_2,n_3)$ = (0,1,0), (0,2,0) and (0,1,1). $n_1$, $n_2$ and $n_3$ are
the OH stretching, bending and OD stretching quantum numbers, respectively. This distribution is represented in the 
top panel of Fig.~\ref{fig:plot6}.

In reference 41, 1 000 000 trajectories were run on the Ochoa-Clary (OC) potential energy surface (PES) \cite{Ochoa}
using the VENUS96 code. 
Initial conditions were selected to reproduce the experiment of Davis and co-workers, 
with a collision energy of 6.6 kcal mol$^{-1}$ and the reactants in their vibrational ground states. 
The number of reactive trajectories was found equal to $N_P = 10837$. At the end of each 
reactive trajectory, the vibrational actions of the triatomic HOD product were calculated using the recent normal mode analysis 
algorithm \cite{Joaquin2}. The latter includes anharmonicity and Coriolis-coupling terms, and yields results similar 
to those obtained by means of the widely used fast Fourier transform approach \cite{Schatz2}, but at a lower computational cost. 

The different distributions previously considered are calculated as follows.
Formally, the purely classical translational energy distribution is given by
Eqs. \eqref{9a} and \eqref{9b} with $N$ equal 3.
\emph{Stricto-sensu}, its Monte-Carlo expression is 
\\
\begin{equation}
P_C(E_T) = \frac{1}{N_T} \sum_{k=1}^{N_T}\;\delta(E_T-E^k_T)
\label{43}
\end{equation}
\\
where $E^k_T$ is the final translational energy for the $k^{th}$ trajectory. This energy satisfies the relation
\\
\begin{equation}
E^k_T=E-\sum_{i=1}^3\;\omega_i (x^k_i+\frac{1}{2})-E^k_R,
\label{44}
\end{equation}
\\
the $x^k_i$'s and $E^k_R$ being the final vibrational actions and 
rotational energy for the $k^{th}$ trajectory.

Here, we shall not replace the Dirac distribution in the previous sum by a Gaussian and calculate it for fixed values
of $E_T$. Instead, we divide the available range of energy $[0, E]$ in $N_r$ boxes $[(i-1)E/N_r,iE/N_r]$,
$i=\overline{1,N_r}$, and integrate $P_C(E_T)$ over the boxes. This leads to the $N_r$ populations
\\
\begin{equation}
P_i = \frac{N_i}{N_T},
\label{45}
\end{equation}
\\
$i=\overline{1,N_r}$, where $N_i$ is the number of trajectories for which the final translational energy 
belongs to the $i^{th}$ box.

We note that applying Eq. \eqref{43} does only require the calculation of $E^k_T$, not of $x^k_1$, $x^k_2$, $x^k_3$ and $E^k_R$.
On the other hand, the four last quantities are necessary for the calculation of $P_{SB}(E_T)$. This distribution is indeed calculated 
in the same way as $P_C(E_T)$, the only difference being that the translational energy for the $k^{th}$ trajectory is now given by
\\
\begin{equation}
E^k_T=E-\sum_{i=1}^3\;\omega_i (\bar{x}^k_i+\frac{1}{2})-E^k_R
\label{46}
\end{equation}
\\
instead of Eq. \eqref{44}. This difference is similar to the one between Eqs. \eqref{9} and \eqref{10}.

For $P_{GB}(E_T)$, $N_i$ is replaced in Eq. \eqref{45} by
\\
\begin{equation}
N^{GB}_i = \sum_{k=1}^{N_i}\;\Pi_{i=1}^{3}\;G(x^k_i-\bar{x}^k_i),
\label{47}
\end{equation}
\\
the sum being performed over the trajectories for which the final translational energy according to Eq. \eqref{44} belongs to the $i^{th}$ box.

For $P_{1GB}(E_T)$, $N_i$ is replaced by 
\\
\begin{equation}
N^{GB}_i = \sum_{k=1}^{N_i}\;G\Big(\frac{\sum_{i=1}^3\omega_i (x^k_i-\bar{x}^k_i)}{\omega}\Big),
\label{47}
\end{equation}
\\
the sum being performed over the same trajectories as above. 

Finally, $P_{SB}(E_T)$, $P_{GB}(E_T)$ and $P_{1GB}(E_T)$ were re-normalized to one. $\epsilon$ was kept at 0.05
for $P_{GB}(E_T)$, and 0.01 for $P_{1GB}(E_T)$. $\omega$ was identified with the largest frequency.
The distributions are displayed in Fig.~\ref{fig:plot6}. We also kept $\omega$ at the sum of the frequencies,
following Czako and Bowman \cite{Cza}, but due to the re-normalization, this left the distribution unchanged.

Contrary to the purely classical distribution which has a bell shape and does not reproduce the 
vibrational structures observed experimentally \cite{Sierra,Joaquin}, the SB, GB and 1GB distributions  
reproduce quite satisfyingly the two structures due to the (0,1,0) and (0,2,0) vibrational states. On the other hand,
the third structure, due to the (0,1,1) state, is strongly underestimated by all the treatments. Comparison with exact quantum
scattering calculations, certainly possible in a near future, will tell if the previous disagreement is due to possible
inaccuracies of the OC-PES or to the present classical descriptions. 

The SB procedure does a good job, though it overestimates the contribution of the small translational 
energies to the (0,2,0) peak. 

The GB distribution of Fig.~\ref{fig:plot6} involves strong fluctuations, contrary to the same density represented in
Fig. 2 of reference 41. The reason is that in the present work, we did not use the smoothing
procedure previously considered \cite{Joaquin} (two Gaussian functions were used to fit the left and right-hand side of 
each vibrational contribution). We did it on purpose, in order to illustrate the fact that with $\sim$ 11000 
reactive trajectories and three vibrational modes, the usual GB procedure generates quite noisy curves. On the other
hand, the 1GB distribution is much better converged and one guesses that it represents the curve one would 
obtain from smoothing the GB curve. 

Owing to the fact that the OH+D$_2$ reaction is a direct process, the present results are
quite encouraging for future applications of the 1GB procedure to polyatomic reactions, do they involve a long-lived
complex or not.

\section{Conclusion}

In the recent years, many processes have been studied by the quasi-classical trajectory method (QCTM) within 
the Gaussian binning (GB) procedure. In most studies, the population of the final product quantum state 
$\bold{n} = (n_1,...,n_N)$, $N$ being the number of quantized degrees of freedom (DOF), was approximated by
\\
\begin{equation}
P_{\bold{n}} = \frac{1}{N_T}\sum_{k=1}^{N_{\bold{n}}}\;\Pi_{i=1}^N\;G(x^k_i-n_i)
\label{48}
\end{equation}
instead of the usual expression
\begin{equation}
P_{\bold{n}} = \frac{N_{\bold{n}}}{N_T}
\label{49}
\end{equation}
\\
used in the standard  binning (SB) procedure (or histogram method).
$N_T$ is the total number of trajectories run, $\bold{x} = (x_1,...,x_N)$ is the final action state,
$N_{\bold{n}}$ is the number of trajectories ending in the product channel with $\bold{x}$ pointing in the
$N$-dimensional unit cube centered on $\bold{n}$ and $G$ is a Gaussian normalized to unity, with a full width
at half maximum of $\sim$ 10 percent.

Since most processes studied so far by GB-QCTM were triatomic reactions, one single vibrational DOF was quantized, 
meaning that the Gaussian product in Eq. \eqref{48} reduced to one term only. As about 10 percent of the total amount of 
reactive trajectories do actually contribute to the product populations, 10 times more trajectories
had to be run for keeping with the same level of convergence of the predictions as compared with SB-QCTM. 

Nowadays, however, more and more processes under scrutiny involve more than one vibrational mode. For instance, 
the reaction OH+D$_2 \longrightarrow$ HOD+D involves three modes while for the reaction F+CH$_4 \longrightarrow$ FH+CH$_3$,
this number is seven. Consequently, GB-QCTM requires one thousand more trajectories than SB-QCTM for the first process
and ten millions more for the second ! It is thus quite clear that as such, Eq. \eqref{48} has no future in the area
of polyatomic reaction dynamics. 

This is why Czak\'o and Bowman \cite{Cza} recently proposed to "quantize" the total vibrational energy
instead of the vibrational actions, introducing the expression
\\
\begin{equation}
P_{\bold{n}} = \frac{1}{N_T}\sum_{k=1}^{N_{\bold{n}}}\;G\Big(\frac{\sum_{i=1}^N \omega_i(x^k_i-n_i)}{\omega}\Big)
\label{50}
\end{equation}
\\
where $\omega_i$ is the $i^{th}$ normal mode frequency and 
\\
\begin{equation}
\omega = \sum_{i=1}^N \omega_i.
\label{51}
\end{equation}
\\
The key feature of this \emph{ad-hoc} quantization as compared to the previous one-Gaussian-for-one-mode
approach is that only one Gaussian function is used whatever the number of vibrational
DOF of the system, a huge amount of computational time being therefore saved. We called it the 1GB procedure.

The conclusions of the present paper are as follows:
\\
\\
1) For a statistical collision involving two product vibrational modes, the 1GB procedure
is strictly equivalent to the GB procedure provided that $\omega$ is identified with the maximum of the
$\omega_i$'s instead of their sum. 
\\
\\
2) For a statistical collision involving three product vibrational modes, the 1GB procedure
is strictly equivalent to the GB procedure provided that $\omega$ is kept at the maximum frequency in part
of the frequency space, and a more complex expression (see Eq. \eqref{34}) in the remaining part. 
\\
\\
3) For the previous processes and a non statistical but sufficiently smooth distribution in the action space, 
the 1GB procedure leads to results in satisfying agreement with the GB ones. 
\\
\\
4) Finding the expression of $\omega$ for any realistic process involving more than 
three product vibrational modes requires heavy mathematical developments that we did not perform. However, 
one may go round this difficulty by re-normalizing the product state populations. In such a case, $\omega$ can be indifferently 
kept at the maximum of the $\omega_i$'s or their sum. Special care should however be taken with processes 
involving a large amount of vibrationally elastic non reactive trajectories, like ion-molecule reactions. 
The methods proposed in reference 23 (leading to Eq. \eqref{40} of the present work) or 34 can then be used. 
\\
\\
5) The 1GB procedure leads to results in good agreement with the GB one for (a) a non statistical test case 
involving three vibrational modes and (b) the prototype four-atom reaction OH+D$_2$ $\longrightarrow$ HOD+D.
\\
\\
In conclusion, the 1GB procedure might be of great interest for future classical simulations of polyatomic chemical
reaction dynamics in the highly quantum mechanical situation where only a few product vibrational states are 
available.

\newpage

\renewcommand{\theequation}{A.\arabic{equation}}

\setcounter{equation}{0}






\renewcommand{\theequation}{B.\arabic{equation}}

\setcounter{equation}{0}

\section*{Appendix}

Consider (i) the $N$-dimensional space $\bold{\Gamma} = (x_1,...,x_N)$, (ii) a given distribution $\rho(\bold{\Gamma})$, normalized to unity, 
of the position in the previous space and (iii) the quantity $Q'$ depending on $\bold{\Gamma}$ according to  
\begin{equation}
Q'= f(\bold{\Gamma}).
\label{a1}
\end{equation}
\\
The probability that $Q'$ is lower than a given value $Q$  is given by
\\
\begin{equation}
\Pi(Q)= \int\;d\bold{\Gamma}\;\rho(\bold{\Gamma})\Theta(Q-Q') = \int\;d\bold{\Gamma}\;\rho(\bold{\Gamma})\Theta(Q-f(\bold{\Gamma})) 
\label{a2}
\end{equation}
\\
where $\Theta(x)$ is the Heaviside function, equal to 0 for $x$ negative and 1 in the contrary case. $\Theta(Q-Q')$ ensures that
integration with respect to $\bold{\Gamma}$ is made over the volume such that $Q$ minus $Q'$ is positive, i.e.,
$Q'$ is lower than $Q$. 

If $P(Q)$ is the density of probability that $Q'$ takes the value $Q$, $P(Q)dQ$ is the probability that $Q'$ 
belongs to the range [$Q$, $Q+dQ$]. We then have 
\\
\begin{equation}
P(Q)dQ = \Pi(Q+dQ)-\Pi(Q)
\label{a3}
\end{equation}
that is,
\begin{equation}
P(Q) = \frac{d\Pi}{dQ}.
\label{a3}
\end{equation}
\\
From Eq. \eqref{a2}, we finally arrive at
\\
\begin{equation}
P(Q)= \int\;d\bold{\Gamma}\;\rho(\bold{\Gamma})\delta(Q-f(\bold{\Gamma})), 
\label{a4}
\end{equation}
\\
as the Dirac distribution $\delta(x)$ is the first derivative of $\Theta(x)$.
Eq. \eqref{9} is straightforwardly obtained from Eqs. \eqref{a4}, \eqref{7} and \eqref{8}.

\section*{Acknowledgments}

LB is endebted to Pr. J.-C. Rayez for careful reading of the manuscript prior to its publication
as well as stimulating discussions on its content and more generally, on the semi-classical description
of molecular collisions.

\newpage

\section*{Figures captions}

Fig. \ref{fig:plot1}: The action space defined by Eq. \eqref{20}
and both $x_1$ and $x_2$ greater than minus 1/2 is represented here by a green triangle while the available quantum states
are represented by six red dots. The distribution of the action pair $(x_1,x_2)$
is uniform in the triangle. Each dot lies along a straight-line corresponding
to a given translational energy (see Eq. \eqref{21}).
Three forbidden quantum states are represented by dark blue dots and
three unit squares centered on quantum states are emphasized. The two salmon ones
correspond to available quantum states while the light blue one corresponds to
a forbidden state. \\

Fig. \ref{fig:plot2}: Translational energy distributions corresponding to the
statistical distribution in the green triangle of Fig. \ref{fig:plot1}. 
The curves are labeled by the subscript of their mathematical
symbols (see text). The distributions are not normalized to unity. \\

Fig. \ref{fig:plot3}: Drawing shedding light on the derivation of 
Eq. \eqref{29d}. $cos\theta$ clearly appears to be the reciprocal of 
the section of the new coordinate axis $y_1$ being within the red square
(see Eq. \eqref{29}).  \\

Fig. \ref{fig:plot4}: Translational energy distributions corresponding to the
statistical distribution in the green triangle of Fig. \ref{fig:plot1}. 
The curves are labeled by the subscript of their mathematical
symbols (see text). The distributions are not normalized to unity. \\

Fig. \ref{fig:plot5}: Translational energy distributions corresponding to the
non statistical distribution given by Eq. \eqref{41}. 
The curves are labeled by the subscript of their mathematical
symbols (see text). The distributions are not normalized to unity. \\

Fig. \ref{fig:plot6}: Translational energy distributions in the products
of the reaction OH+D$_2$ $\longrightarrow$ HOD+D studied at the conditions
of the group of Davis \cite{Davis}. The top curve is the
experimental distribution while the remaining curves are labeled by 
the subscript of their mathematical symbols (see text).
The distributions are normalized to unity. \\

\newpage

\section*{Figures}

\end{document}